\documentclass[fleqn,usenatbib]{mnras}
\usepackage[T1]{fontenc}
\usepackage{ae,aecompl}

\usepackage{graphicx}	% Including figure files
\usepackage{amsmath}	% Advanced maths commands
\usepackage{amssymb}	% Extra maths symbols
\usepackage{multirow}
\usepackage{soul}
\usepackage{color,colortbl}
\usepackage{subfig}
\usepackage{txfontsb}
\usepackage{upgreek}
\usepackage{glossaries}
\glsdisablehyper

\newcommand{\ud}{\rmn{d}}
\newcommand{\kB}{k_{\rmn{B}}}
\newcommand{\TX}{T_{\rmn{X}}}
\newcommand{\dotTX}{\dot T_{\rmn{X}}}
\newcommand{\kT}{\kB\TX}
\DeclareMathOperator{\erf}{erf}

\newacronym{mcmc}{MCMC}{Markov Chain Monte Carlo}
\newacronym{dfe}{DfE}{departure from equilibrium}
\newacronym{ovr}{OVR}{observed virial ratio}
\newacronym{evr}{EVR}{equilibrium virial ratio}
\newacronym{dm}{DM}{dark matter}
\newacronym{de}{DE}{dark energy}
\newacronym{flrw}{FLRW}{Friedmann-Lema\^itre-Robertson-Walker}

\title[Constraints on DE-DM interaction with clusters]{New observational constraints on interacting dark energy using galaxy clusters virial equilibrium states}
\author[M. Le~Delliou et al.]{
    M. Le~Delliou,$^{1,2,3}$\thanks{E-mail: delliou@ift.unesp.br}
    R. J. F. Marcondes$^{4}$\thanks{E-mail: rafaelmarcondes@usp.br} and 
    G. B. Lima~Neto$^{5}$
\\
$^{1}$Institute of Theoretical Physics,Physics Department, Lanzhou University, No.222, South Tianshui Road, Lanzhou, Gansu 730000, P R China\\
$^{2}$Instituto de Astrof\'isica e Ci\^encias do Espa\c co, Universidade de Lisboa, Faculdade de Ci\^encias, Ed. C8, Campo Grande, 1769-016 Lisboa, Portugal\\
$^{3}$Instituto de F\'isica Te\'orica, Universidade Estadual de S\~ao Paulo (IFT-UNESP), Rua Dr. Bento Teobaldo Ferraz 271, Bloco 2 - Barra Funda, \\ CEP 01140-070, S\~ao Paulo, SP, Brazil\\
$^{4}$Instituto de F\'isica, Universidade de S\~ao Paulo, Rua do Mat\~ao 1371, Cidade Universit\'aria, CEP 05508-090, S\~ao Paulo, SP, Brazil\\
$^{5}$Instituto de Astronomia, Geof\'isica e Ci\^encias Atmosf\'ericas, Universidade de S\~ao Paulo, Rua do Mat\~ao 1226, Cidade Universit\'aria, \\ CEP 05508-090, S\~ao Paulo, SP, Brazil
}

\date{Accepted XXX. Received YYY; in original form ZZZ}

\pubyear{2018}

\begin{document}
\label{firstpage}
\pagerange{\pageref{firstpage}--\pageref{lastpage}}
\maketitle

\begin{abstract}
    As the dark sector remains unknown in composition and
    interaction between dark energy and dark matter stand out as
    natural, observations of galaxy clusters out of
    equilibrium abound, opening a promising window on these
    questions. We continue here the exploration of dark
    sector interaction detection via clusters virial
    equilibrium state for all clusters configurations. The
    dynamics of clusters is evaluated with the Layzer-Irvine
    equation, a simple model of an interacting dark sector and
    some simplifying assumptions to obtain the
    time-dependent part of the virial dynamics. The
    clusters' data are concentrated in optical weak lensing
    and X-ray observations that evaluate, respectively, the
    clusters' mass profiles and temperatures. The global
    inconsistency of available X-ray data led us to
    constitute ``gold'' cluster samples. Through a Bayesian
    analysis, they are processed to obtain consistent
    interaction detected up to 3$\sigma$, in compounded
    interaction strength for 11 clusters at
    $-0.027 \pm 0.009$ that translate in compounded
    universal equilibrium virial ratio of
    $-0.61^{+0.04}_{-0.03}$. The level of detection and
    inconsistency of X-ray data call for caution, although
    future instruments promise a clearer detection soon.
\end{abstract}


%%%%%%%%%%%%%%%%%%%%%%%%%%%%%%%%%%%%%%%%%%%%%%%%%%

%%%%%%%%%%%%%%%%% BODY OF PAPER %%%%%%%%%%%%%%%%%%

\section{Introduction}
Since the discovery of cosmic acceleration \citep{perlmutter_measurements_1999}
and the proposal of \gls{de} as its source, in addition to the already
sought \glsentrylong{dm}
\citep[\glsentryshort{dm},\glsunset{dm}][]{zwicky_rotverschiebung_1933,
    zwicky_masses_1937}, the largely unknown nature of the dark sector naturally
called for possible interactions within its manifestations
\citep{amendola_coupled_2000, amendola_perturbations_2000}.

Despite considerable efforts towards direct and indirect detection, the only
evidence at hand of the existence of the dark sector remain purely
gravitational, through the Cosmic Microwave Background observations
\citep{ade_planck_2014}, supernovae acceleration
\citep{perlmutter_measurements_1999, riess_observational_1998} or clusters
displaying segregated mass and baryons (dissociative
clusters), such as the so-called 
``Bullet Cluster'' \citep{clowe_direct_2006},
``El Gordo'' \citep{jee_weighing_2014},
Abell~1758 \citep{monteiro-oliveira_merger_2017},
among others.
In this context, detection of interactions inside the dark
sector would significantly help us understand the nature of
dark matter and dark energy and even increase the
probability of these components to exist.

In a previous paper \citep[][hereafter LeD15]{le_delliou_non-virialized_2015},
we developed an approach to the detection of such interactions in the virial
state of galaxy clusters, through a simplified coupled dark energy (CDE)
cosmology model, coupled with the Layzer-Irvine dynamical virial equation. Based
on a series of papers exploring such detection in apparently balanced clusters,
and their check by other groups \citep{bertolami_dark_2007,
    le_delliou_dark_2007, bertolami_dark_2008, bertolami_abell_2009,
    bertolami_testing_2012, abdalla_signature_2009, abdalla_signature_2010,
    he_imprint_2010}, this latest approach attempted to include the effect of
departure from equilibrium.
However, although this allowed for the use in the detection of a wider sample of
clusters, it involved the assumption that clusters present
small departures from virial equilibrium, and found it to be the source of
inconsistencies in the results. The present paper proposes now to remedy these
inconsistencies by allowing
larger departures in an evaluation independent from the astrophysical processes
expected to
source this deviation from balance.
We also attempt a more robust statistical treatment of the data, with a Bayesian
approach.

In the following section \ref{sec:Framework}, the framework in which data will
be analysed is laid out. The sample and statistical treatment are discussed in
Sec.~\ref{sec:data}, while the analysis is described in Sec.~\ref{sec:MCMC}. The
results are discussed in Sec.~\ref{sec:Results} before to conclude in
Sec.~\ref{sec:conclusions}.

\section{The Framework}
\label{sec:Framework}
\subsection{The cosmological model}
We model the universe, composed of dark matter and dark energy
only, as a flat \gls{flrw} background metric.
The dark sector interaction is modeled with a heat flux in the Bianchi
identities between the two dark components, denoted by subscript $c$ for cold
\gls{dm} and $d$ for \gls{de} (i.e. energy conservation equations, linking the
energy densities $\rho$ evolutions, the \gls{de} equation of state
$w_d=P_d/\rho_d$, $P_d$ being the \gls{de} pressure, and the Hubble parameter
$H$ to the dark matter-dark energy interaction coupling $\xi$):
\begin{equation}
	\label{eq:Bianchi}
    \dot\rho_c + 3 H \rho_c = 3 H \xi \rho_c, \qquad
    \dot\rho_d + 3 H \rho_d \left(1 + w_d \right) = - 3 H \xi \rho_c.
\end{equation}
With this sign convention, positive $\xi$ means that dark energy decays
into dark matter.
The equation-of-state parameter $w_d$ is set to $-1$ in most of our analyses, except in one case where we make it a free parameter of the model.
The rest of the \gls{flrw} evolution is standard.

\subsection{The Layzer-Irvine equation}
\label{ss:LIequation}
The Layzer-Irvine equation can be recast to relate the kinetic ($\rho_K$) and
gravitational potential ($\rho_W$) parts of the dark matter density $\rho_c$ of
the studied, evolving system (a cluster). As a generalisation of the virial
equation, it describes how the system tends to relax. In this CDE scenario, it
has been obtained by \cite{he_imprint_2010} as
\begin{equation}
    \label{eq:LIp2}
    \dot\rho_c + H \left[ \left(2 + 3 \xi \right) \rho_K + \left(1 - 6\xi\right) \rho_W \right] = 0.
\end{equation}
In LeD15,\footnote{
    In LeD15, the choice of the coupling strengths $\xi_1 = \xi/18$ and $\xi_2 =
    -(\xi/6) \, \rho_{c}/\rho_{d}$ was inconsistent with the derivation of the
    Layzer-Irvine equation by \cite{he_imprint_2010}, which defines $\xi_1$ and
    $\xi_2$ to be constants.
    Here, we amend that mistake simply adopting $\xi_1 = \xi$ and $\xi_2 = 0$,
    which also makes the interaction dependent on the dark matter energy density
    only, but leads to a different Layzer-Irvine equation.
    Notice that the sign of the interacting term yields a positive flux $3H\xi
    \rho_c$ towards \gls{dm} when all terms are positive, in agreement with common
    phenomenological descriptions of the interacting term in the literature
    \citep[for instance][]{he_imprint_2010, cao_interaction_2013,
        costa_constraints_2017}.} 
the condition of small departures from equilibrium was imposed, that led to the
approximation $\dot\rho_K/\rho_K\simeq\dot\rho_W/\rho_W$.
In this work, the results from LeD15 require to allow the clusters to be away
from equilibrium.
Thus $\dot\rho_W$ and $\dot\rho_K$ will be modeled separately (see
Sec.~\ref{sss:DfE}).
Eq.~(\ref{eq:LIp2}) can be reformulated to give the out-of-equilibrium virial
ratio
\begin{equation}
    \label{eq:VRLI}
    \frac{\rho_K}{\rho_W} = - \frac{1 - 6\xi}{2 + 3\xi} - \frac{1}{2 + 3\xi}
    \frac{\dot \rho_K +  \dot\rho_W}{H \rho_W}.
\end{equation}

This allows us to compare observed values of the virial
ratio, built from the quantity $\rho_K/\rho_W$ extracted
from clusters and called hereafter the \gls{ovr}, with a
modified ratio involving the interaction coupling, which we
will refer to as the \gls{evr},\footnote{Formerly named
    theoretical virial ratio (TVR) in LeD15.} and the time
evolution term involving the time derivative, which we call
\gls{dfe}.
We propose to model and build the \gls{ovr} and \gls{dfe} from observations of
clusters' mass $M_{200}$ enclosed in a radius
$r_{200}$,\footnote{\label{fn:r200eval}Recall then that
    $M_{200}/\rmn{Vol}(r_{200})=200\,\rho_{cr}$ with the critical density
    $\rho_{cr} = 3H^2(z)/8\uppi G$.} the NFW concentration parameter $c_{200}$
and the X-ray temperature $\TX$.
The \gls{dfe} will also depend on the parameter of interest $\xi$, on the
density parameter $\Omega_{c0}$ and on $h \equiv H_0 / 100 \, \rmn{km} \,
\rmn{s}^{-1} \, \rmn{Mpc}^{-1}$ that enter in the Hubble rate $H$.
Rewriting Eq.~(\ref{eq:VRLI}) as
\begin{equation}
    \label{eq:LI}
    \frac{\rho_K}{\rho_W} + \frac{1}{2 + 3 \xi} \frac{\dot\rho_K + \dot \rho_W}{H \rho_W} = - \frac{1 - 6\xi}{2 + 3\xi},
\end{equation}
explicit the universal, predicted equilibrium virial ratio (i.e., the kinetic to
potential ratio that should be reached by a cluster at perfect equilibrium) that
can be obtained from specific clusters' observed virial ratio minus departure
from equilibrium in the left-hand side of Eq.~(\ref{eq:LI}).
The first step is to evaluate the kinetic and potential energy densities. Then
we need to evaluate in a sensible way the \gls{dfe} term.
Thus only remains to place constraints on the interaction coupling parameter
$\xi$, which can be performed by \gls{mcmc} simulations.

\subsubsection{The kinetic and potential energy densities}

We follow LeD15 evaluations of these densities, from the measurements of the
given cluster's X-ray temperature $\TX$, mass $M_{200}$ and NFW concentration
parameter $c_{200}$. The potential energy is approximated using the NFW density
profile \citep{navarro_structure_1996} extracted from the cluster's observed
mass and concentration  (defining $c_{200} = r_{200}/r_0$ instead of using
$r_0$).
Thus, we have
\begin{equation}
    \rho_W = - \frac{3 G M_{200}^2}{4\uppi r^4_{200} f_c}, 
\end{equation}
with
\begin{equation}
    f_{c}
    \equiv\frac{C^2/c_{200}}{\frac{1}{2} c_{200}^2 - C}, 
    \quad C \equiv C' \ln C' - c_{200},
    \quad C' \equiv 1+c_{200}.
\end{equation}
The kinetic energy is (LeD15)
\begin{equation}
    \rho_K = \frac{9}{8\uppi} \frac{M_{200}}{r_{200}^3} \frac{\kT}{\mu m_H},
\end{equation}
where $\kB$ is the Boltzmann constant, $\mu = 0.63$ is the
intracluster plasma mean molecular weight (defined as the
mean mass of the particles divided by the Hydrogen mass,
assumed to be completely ionized and with primordial
chemical composition), and $m_H$ is the proton mass.
The ratio of these two densities is the observed virial ratio
\begin{equation}
    \label{eq:OVR}
    \frac{\rho_K}{\rho_W} = - \frac{3}{2} \frac{r_{200}}{G M_{200}} \frac{\kT}{\mu m_H} f_c.
\end{equation}
The radius $r_{200}$ is evaluated from the NFW parameters (see
footnote~\ref{fn:r200eval}) with the critical density at the
redshift of the cluster and in the same cosmology assumed by the observers
to keep consistency with the fitted NFW profile.

\subsubsection{Evaluating the departure from equilibrium}
\label{sss:DfE}
To allow for the extra freedom introduced by relaxing the small departures from
equilibrium assumption, compared to LeD15, the virial ratio now depends on both
temperature and virial radius, the concentration remaining a parameter.
We note that both densities can be rewritten as functions of their local measured
quantities, recognizing the critical density ratio definition in powers of mass
and radius, as
\begin{equation}
    \rho_K = \rho_K(\TX) = 300 \, \rho_{cr} \frac{\kT}{\mu m_H}
\end{equation}
and
\begin{equation}
    \rho_W = \rho_W(r_{200}) = - \left(200 \, \rho_{cr} \right)^2 G \frac{4
        \uppi}{3} \frac{r_{200}^2}{f_c}.
\end{equation}
We thus can compute the time derivatives of Eq.~(\ref{eq:VRLI}) as
\begin{equation}
    \dot\rho_K + \dot\rho_W = \frac{\ud \rho_K}{\ud \TX} \dotTX + \frac{\ud \rho_W}{\ud r_{200}} \dot r_{200},
\end{equation}
which is fully general, as opposed to the evaluation in LeD15. The delicate part is then to evaluate $\dotTX$ and $\dot r_{200}$.
Based on the reasonable expectations from hierarchical structure formation that
clusters' temperature and radius should
evolve to equilibrium values, increasing faster in the past than in
the future, we propose two physically reasonable ansatze
which  derivatives asymptote to zero from positive
decreasing values, meaning that both $\TX$
and $r_{200}$ should increase to reach equilibrium, a
behaviour which is observed in semi analytical simulations
\citep[as can be seen from
studying][]{henriksen_stationary_1995,
    henriksen_self-similar_1997, henriksen_relaxing_1999,
    del_popolo_density_2000, le_delliou_non-radial_2003,
    macmillan_universal_2006, le_delliou_merger_2008}:
\begin{equation}
    \dotTX = \frac{\TX/t_0}{(t/t_0)^2}, \quad \dot r_{200} = \gamma \frac{r_{200}/t_0}{(t/t_0)^{\gamma+1}},
\end{equation}
derived respectively from heuristic exponential parametrizations $\TX = \TX^*
\exp(-t_0/t)$ and $r_{200} = r_{200}^* \exp(-t_0/t)^{\gamma}$ ($\gamma \ne 1$),
where $\TX^*$ and $r_{200}^*$ are the asymptotic equilibrium
values and $t_0$ is some characteristic time scale.
The parametrization is using the simplicity of the strong convergence of the
exponential function \citep[see, e.g. the fast virialization of haloes seen
in][and their moderately violent relaxation]{henriksen_stationary_1995} and the
finite value convergence of the inverse power law.
We further restrict our parametrization of $\gamma$ to positive values so as to
keep the approach of asymptotic growth of the radius towards $r_{200}^{*}$. 
When $\gamma < 1$ ($\gamma > 1$), the radius approaches the
equilibrium faster (slower) than the
temperature.\footnote{The two cases are better analysed
    separately due to divergences at $\gamma = 1$. For the
    sake of simplicity, in this work we consider only the
    first case.}
These ansatze are used locally to give the evolution slopes but
are not considered globally integrable.
They provide one equation,
\begin{equation}
    \frac{\dotTX}{\TX /t_0} =  \left(\frac{\dot r_{200}}{\gamma r_{200}/t_0}\right)^{\frac{2}{\gamma +1}},
\end{equation}
to obtain the unknown time evolutions $\dotTX$ and $\dot r_{200}$. The remaining equation needed to provide a solution in terms of observed values for these unknown can be chosen as the equation of state for the perfect gas, considered isobaric:
\begin{equation}
    \frac{T_{X}}{r_{200}^{3}}= \rmn{constant},
\end{equation}
which can be derived into
\begin{equation}
    \frac{\dotTX}{\dot r_{200}} = 3 \frac{\TX}{r_{200}}.
\end{equation}
Solving for the derivatives in terms of $\gamma$, $\TX$ and $r_{200}$, the
\gls{dfe} term is given by
\begin{equation}
    - \frac{\dot\rho_K + \dot\rho_W}{\left(2 + 3\xi\right) H \rho_W} = 
    - \left( 3 \frac{\TX}{\rho_W} \frac{\ud \rho_K}{\ud \TX} + \frac{r_{200}}{\rho_W}
        \frac{\ud \rho_W}{\ud r_{200}} \right)
    \frac{ \left( \gamma^2 3^{\gamma+1}\right)^{\frac{1}{1-\gamma}}}{\left(2 + 3\xi\right) H t_0},
\end{equation}
with the derivatives given by
\begin{equation}
    \frac{\TX}{\rho_W} \frac{\ud \rho_K}{\ud \TX} =
    \frac{\rho_K}{\rho_W} \frac{\ud \ln \rho_K}{\ud \ln \TX} = \frac{\rho_K}{\rho_W}
\end{equation}
and 
\begin{equation}
    \frac{r_{200}}{\rho_W} \frac{\ud \rho_W}{\ud r_{200}} = 2 - \frac{\ud \ln
        f_c}{\ud \ln r_{200}} = 2 - \frac{\ud \ln f_c}{\ud \ln c_{200}}.
\end{equation}
The exact time scale $t_0$ is not important to our
purposes. Since this parameter only appears dividing
$(\gamma^2 3^{\gamma+1})^{1/(1-\gamma)}$, it can
be absorbed into this term with the only effect of shifting
the value of $\gamma$ at which its marginalized distribution
becomes suppressed (as that term diverges with $\gamma$
approaching the unity), so we set $t_0 = 1$ (in units of $\rmn{km}^{-1} \, \rmn{s} \, \rmn{Mpc}$).

\section{The data}
We start from a sample of 50 clusters with weak-lensing mass measurements of
$M_{200}$ given by \citet{okabe_locuss:_2016} and corresponding measurements of
$c_{200}$ kindly provided by Okabe (private communication).
The NFW profiles are based on a flat $\Lambda$CDM background cosmology with
\gls{dm} and \gls{de} density parameters $\Omega_{c0} = 0.3$
and $\Omega_{d0} = 0.7$, which we use in the evaluation of $r_{200}$.
These data can be complemented with X-ray temperature data from a few different
sources.
By collecting temperature data from 
\citet{maughan_self-similar_2012}, \citet{martino_locuss:_2014} or
\citet{mantz_weighing_2016, mantz_erratum:_2017} (hereafter M12, M14 and M16,
respectively), we end up with subsets of 22, 19 or 30 clusters.
The data are summarized in table~\ref{tab:clusters_data}.
\begin{table*}
    \caption{Redshift, NFW parameters from \citet{okabe_locuss:_2016} and
        temperature of galaxy clusters from different sources.
        Temperatures are given in $\rmn{keV}$ and masses in units of
        $h^{-1}10^{14}M_{\odot}$.}
    \label{tab:clusters_data} %
    \renewcommand{\arraystretch}{1.4}
    \begin{tabular}{@{}lcccccc@{}}
        \hline  % 150414-2342,2352,2358 2600
        Cluster  & $z$  &  $M_{200}$    &   $c_{200}$   & 
        $\kT$ \citep{maughan_self-similar_2012} & 
        $\kT$ \citep{martino_locuss:_2014} & 
        $\kT$ \citep{mantz_weighing_2016, mantz_erratum:_2017} \\
        \hline 
        ABELL0068  &  $0.2546$  &  $6.65^{+1.35}_{-1.16}$  & $4.83^{+1.83}_{-1.31}$  &  $7.8 \pm 1.0$ & $5.02 \pm 1.65$  &  $9.62 \pm 1.65$    \\
        ABELL0115  &  $0.1971$  &  $7.04^{+2.66}_{-1.97}$  & $1.59^{+1.12}_{-0.77}$  &  $6.7 \pm 0.3$ & $6.46 \pm 0.51$  &  $11.74 \pm 0.90$   \\
        ABELL0209  &  $0.2060$  &  $12.75^{+2.27}_{-1.91}$  & $3.63^{+1.02}_{-0.84}$  &  $7.4 \pm 0.5$ & $7.56 \pm 1.40$  &  $8.98 \pm 0.67$   \\
        ABELL0267  &  $0.2300$  &  $5.96^{+1.16}_{-1.08}$   & $3.16^{+1.01}_{-0.81}$  &  $4.4^{+0.5}_{-0.4}$ &        --        &         --         \\
        ABELL0383  &  $0.1883$  &  $5.23^{+1.30}_{-1.07}$  & $4.12^{+2.06}_{-1.41}$  &  $4.5 \pm 0.3$ & $5.76 \pm 1.26$  &  $7.26 \pm 0.42$    \\
        ABELL0521  &  $0.2475$  &  $5.61^{+1.18}_{-1.05}$  &  $3.48^{+1.57}_{-1.09}$  &  $4.8 \pm 0.2$ &       --      &  $7.29 \pm 0.25$  \\
        ABELL0586  &  $0.1710$  &  $6.65^{+2.15}_{-1.61}$  &  $6.77^{+6.83}_{-3.36}$  &  $7.6 \pm 0.8$ &       --      &  $7.40 \pm 0.53$  \\
        ABELL0697  &  $0.2820$  &  $9.74^{+2.90}_{-2.13}$  &  $1.75^{+1.00}_{-0.75}$  &  $10.2^{+0.8}_{-0.7}$ &       --     &  $14.58 \pm 1.44$  \\
        ABELL0750  &  $0.1630$  &  $6.30^{+2.71}_{-1.74}$  &  $3.79^{+2.72}_{-1.68}$  &       --     &       --       &  $6.04 \pm 0.38$  \\
        ABELL0773  &  $0.2170$  &  $9.56^{+1.28}_{-1.14}$  & $5.67^{+1.58}_{-1.27}$  &  $7.4 \pm 0.4$ & $8.64 \pm 2.05$  &  $8.97 \pm 0.52$    \\
        ABELL0781  &  $0.2984$  &  $6.57^{+1.97}_{-1.65}$  & $2.32^{+2.16}_{-1.32}$  &  $5.5^{+0.7}_{-0.5}$ & $5.64 \pm 2.22$  &         --          \\
        ABELL0907  &  $0.1669$  &  $14.28^{+4.59}_{-2.99}$  &  $1.86^{+0.94}_{-0.72}$  &  $5.4 \pm 0.2$ & $6.23 \pm 0.45$  &  $7.17 \pm 0.26$  \\
        ABELL0963  &  $0.2050$  &  $7.13^{+1.38}_{-1.20}$  &  $3.77^{+1.38}_{-1.05}$  &       --     &       --       &  $7.60 \pm 0.37$  \\
        ABELL1423  &  $0.2130$  &  $4.30^{+1.19}_{-0.97}$  &  $5.03^{+4.17}_{-2.30}$  &       --     &       --       &  $7.04 \pm 0.45$  \\
        ABELL1682  &  $0.2260$  &  $8.66^{+1.38}_{-1.21}$  &  $3.93^{+1.00}_{-0.83}$  &  $5.8^{+2.0}_{-1.2}$ &       --      &  $7.67 \pm 0.74$  \\
        ABELL1689  &  $0.1832$  &  $10.98^{+1.66}_{-1.46}$  &  $10.56^{+4.31}_{-2.81}$  &  $8.4^{+0.4}_{-0.3}$ &   $11.23 \pm 1.06$   &  $10.92 \pm 0.32$ \\
        ABELL1763  &  $0.2279$  &  $16.92^{+3.42}_{-2.70}$  &  $3.11^{+1.09}_{-0.86}$  &  $8.1 \pm 0.5$ &   $7.98 \pm 1.45$  &  $9.09 \pm 0.67$   \\
        ABELL1835  &  $0.2528$  &  $10.09^{+1.88}_{-1.63}$  &  $6.94^{+4.29}_{-2.35}$  &       --     &   $11.06 \pm 1.09$  &  $12.15 \pm 0.45$  \\
        ABELL1914  &  $0.1712$  &  $8.73^{+1.92}_{-1.59}$  &  $2.64^{+1.03}_{-0.81}$  &  $8.5^{+0.6}_{-0.4}$ &   $8.57 \pm 1.57$  &  $9.67 \pm 0.50$    \\
        ABELL2009  &  $0.1530$  &  $7.78^{+3.19}_{-2.03}$  &  $1.96^{+1.61}_{-0.96}$  &       --     &       --       &  $7.37 \pm 0.47$  \\
        ABELL2111  &  $0.2290$  &  $4.93^{+2.68}_{-1.48}$  &  $4.98^{+9.01}_{-3.92}$  &   $6.4^{+0.7}_{-0.6}$ &       --      &  $9.07 \pm 0.70$  \\
        ABELL2204  &  $0.1524$  &  $9.56^{+2.29}_{-1.83}$  &  $5.17^{+2.10}_{-1.48}$  &  $8.4^{+0.8}_{-0.6}$ &    $10.50 \pm 1.11$  &  $14.98 \pm 0.72$  \\
        ABELL2219  &  $0.2281$  &  $10.40^{+2.15}_{-1.75}$  &  $2.04^{+0.99}_{-0.75}$  &       --     &       --       &  $12.80 \pm 0.36$  \\
        ABELL2261  &  $0.2240$  &  $11.93^{+2.18}_{-1.80}$  &  $2.69^{+0.93}_{-0.74}$  &  $7.3 \pm 0.4$ &       --      &  $8.75 \pm 0.49$  \\
        ABELL2390  &  $0.2329$  &  $10.60^{+1.91}_{-1.67}$  &  $4.11^{+1.15}_{-0.96}$  &       --     &       --       &  $15.47 \pm 0.68$  \\
        ABELL2485  &  $0.2472$  &  $5.72^{+1.33}_{-1.13}$  &  $3.44^{+2.11}_{-1.34}$  &       --     &       --       &  $7.37 \pm 0.94$  \\
        ABELL2537  &  $0.2966$  &  $7.65^{+2.31}_{-1.90}$  &  $8.76^{+10.55}_{-4.29}$  &       --     &    $8.68 \pm 3.78$   &  $9.22 \pm 0.61$  \\
        ABELL2552  &  $0.2998$  &  $7.61^{+2.88}_{-2.08}$  &  $3.20^{+3.16}_{-1.76}$  &       --     &       --       &  $10.43 \pm 1.34$  \\
        ABELL2631  &  $0.2779$  &  $7.13^{+2.07}_{-1.66}$  &  $1.73^{+2.10}_{-1.03}$  &  $6.9^{+0.8}_{-0.5}$ &    $6.50 \pm 1.20$  &        --          \\
        ABELL2645  &  $0.2510$  &  $4.16^{+1.15}_{-0.99}$  &  $3.58^{+2.30}_{-1.39}$  &       --       &       --       &  $7.30 \pm 1.53$  \\
        ABELL2813  &  $0.2924$  &  $8.17^{+1.91}_{-1.61}$  &  $4.99^{+2.96}_{-1.83}$  &       --     &    $5.48 \pm 1.13$   &       --          \\
        RXJ1504.1$-$0248  &  $0.2153$  &  $5.53^{+1.46}_{-1.25}$  &  $14.75^{+11.69}_{-5.44}$  &  $9.4^{+1.1}_{-1.0}$ &       --     &  $15.31 \pm 1.09$  \\
        RXJ1720.1+2638  &  $0.1640$  &  $5.23^{+1.96}_{-1.45}$  &  $3.27^{+1.96}_{-1.31}$  &  $6.8^{+0.5}_{-0.3}$ & $7.46 \pm 1.03$  &  $9.45 \pm 0.48$ \\
        RXJ2129.6+0005  &  $0.2350$  &  $4.69^{+1.63}_{-1.29}$  &  $1.44^{+1.51}_{-0.86}$  &  $6.2 \pm 0.6$ & $7.62 \pm 1.35$  &  $8.31 \pm 0.44$ \\
        ZwCl1021.0+0426 &  $0.2906$  &  $5.24^{+1.09}_{-0.96}$  &  $4.60^{+2.32}_{-1.52}$  &       --     & $10.48 \pm 2.10$  &        --          \\
        ZwCl1459.4+4240 &  $0.2897$  &  $8.54^{+1.22}_{-1.08}$  &  $3.91^{+0.92}_{-0.77}$  &       --     & $6.41 \pm 2.76$  &        --          \\
        \hline 
    \end{tabular}
\end{table*}

We note that uncertainties in $M_{200}$, $c_{200}$ and $\kT$ from
M12 are generally asymmetrical, in
the form $\bar x^{+\sigma_{+}}_{-\sigma_{-}}$.
Since these quantities should be always positive and typically $\sigma_{+} \ge
\sigma_{-}$, it seems reasonable to assume that these $1\sigma$-error measurements
represent well 68.3 per cent credible intervals of lognormal distributions.
As in LeD15, we want these lognormal distributions to have their parameters
$\mu$ and $\sigma$ adjusted to match the following conditions: 
(i) the maximum probability coincides with the nominal value $\bar x$,
(ii) the probability of the random variable lying between $\bar x - \sigma_{-}$
and $\bar x + \sigma_{+}$ is 68.3 per cent and
(iii) the probability density function has the same value at the points $\bar x
- \sigma_{-}$ and $\bar x + \sigma_{+}$, so that the interval between them
corresponds to the 68.3 per cent most likely values.
For this, we write
\begin{equation}
    \chi^2 = C_{\rmn{(i)}}^2 + C_{\rmn{(ii)}}^2 + C_{\rmn{(iii)}}^2,
\end{equation}
where
%\begin{equation}
%    C_{\rmn{(i)}} \equiv \frac{e^{\mu - \sigma^2}}{\bar x} - 1, 
%    C_{\rmn{(ii)}} \equiv \frac{f_{\mu,\sigma}(\bar x + \sigma_{+})}{f_{\mu,\sigma}(\bar x - \sigma_{-})} - 1, 
%    C_{\rmn{(iii)}} \equiv \frac{F_{\mu,\sigma}(\bar x + \sigma_{+}) - F_{\mu,\sigma}(\bar x - \sigma_{-})}{0.683} - 1
%\end{equation}
\begin{align}
    C_{\rmn{(i)}} &\equiv \frac{\exp\left(\mu - \sigma^2\right)}{\bar x} - 1,  \\
    C_{\rmn{(ii)}} &\equiv \frac{f_{\mu,\sigma}(\bar x + \sigma_{+})}{f_{\mu,\sigma}(\bar x - \sigma_{-})} - 1, \\
    C_{\rmn{(iii)}} &\equiv \frac{F_{\mu,\sigma}(\bar x + \sigma_{+}) - F_{\mu,\sigma}(\bar x - \sigma_{-})}{0.683} - 1
\end{align}
represent the three conditions, with 
\begin{equation}
    f_{\mu,\sigma}(x) = \frac{1}{x \sigma \sqrt{2 \uppi}} \exp\left[- \frac{\left(\ln x - \mu\right)^2}{2 \sigma^2} \right]
\end{equation}
and 
\begin{equation}
    F_{\mu,\sigma}(x) = \frac{1}{2} \left[1 + \erf \left( \frac{\ln x - \mu}{\sigma \sqrt{2}} \right) \right]
\end{equation}
the lognormal probability (PDF) and cumulative (CDF) density functions,
respectively.
We then find, for each of these measurements, the pair of
parameters $(\mu, \sigma)$ that minimizes $\chi^2$.
The parametrizations obtained are verified to match the three conditions
remarkably well in all cases.
We list all the fitted parameters $(\mu, \sigma)$ in
table~\ref{tab:lognormal_parameters}.
\begin{table}
    \caption{Lognormal parameters $(\mu, \sigma)$ for measurements of masses
        (in units of $h^{-1}10^{14}M_{\odot}$), concentrations and temperatures
        \citep[from][in $\rmn{keV}$]{maughan_self-similar_2012} with asymmetrical uncertainties.}
    \label{tab:lognormal_parameters} %
    \renewcommand{\arraystretch}{1.4}
    \begin{tabular}{@{}lccc@{}}
        Cluster  &  $M_{200}$    &   $c_{200}$   & $\kT$ \\ 
        \hline
        ABELL0068 &  (1.93, 0.19)  &  (1.67, 0.30)  &  (2.06, 0.13) \\
        ABELL0115 &  (2.04, 0.31)  &  (0.68, 0.52)  &  (1.90, 0.04) \\
        ABELL0209 &  (2.57, 0.16)  &  (1.34, 0.25)  &  (2.00, 0.07) \\
        ABELL0267 &  (1.81, 0.19)  &  (1.22, 0.28)  &  (1.50, 0.10) \\
        ABELL0383 &  (1.70, 0.22)  &  (1.56, 0.38)  &  (1.51, 0.07) \\
        ABELL0521 &  (1.76, 0.20)  &  (1.37, 0.35)  &  (1.57, 0.04) \\
        ABELL0586 &  (1.97, 0.27)  &  (2.26, 0.59)  &  (2.03, 0.11) \\
        ABELL0697 &  (2.34, 0.25)  &  (0.72, 0.46)  &  (2.33, 0.07) \\
        ABELL0750 &  (1.96, 0.32)  &  (1.56, 0.50)  &  --       \\
        ABELL0773 &  (2.27, 0.13)  &  (1.79, 0.24)  &  (2.00, 0.05) \\
        ABELL0781 &  (1.94, 0.27)  &  (1.14, 0.62)  &  (1.73, 0.11) \\
        ABELL0907 &  (2.74, 0.25)  &  (0.75, 0.41)  &  (1.69, 0.04) \\
        ABELL0963 &  (1.99, 0.18)  &  (1.41, 0.30)  &  --       \\
        ABELL1423 &  (1.51, 0.24)  &  (1.89, 0.53)  &  --       \\
        ABELL1682 &  (2.18, 0.15)  &  (1.41, 0.23)  &  (1.85, 0.26) \\
        ABELL1689 &  (2.42, 0.14)  &  (2.47, 0.31)  &  (2.14, 0.04) \\
        ABELL1763 &  (2.86, 0.18)  &  (1.21, 0.30)  &  (2.09, 0.06) \\
        ABELL1835 &  (2.34, 0.17)  &  (2.14, 0.41)  &  --       \\
        ABELL1914 &  (2.20, 0.20)  &  (1.06, 0.33)  &  (2.15, 0.06) \\
        ABELL2009 &  (2.17, 0.31)  &  (0.94, 0.55)  &  --       \\
        ABELL2111 &  (1.77, 0.37)  &  (2.19, 0.91)  &  (1.87, 0.10) \\
        ABELL2204 &  (2.30, 0.21)  &  (1.75, 0.32)  &  (2.14, 0.08) \\
        ABELL2219 &  (2.38, 0.18)  &  (0.84, 0.40)  &  --       \\
        ABELL2261 &  (2.51, 0.16)  &  (1.07, 0.29)  &  (1.99, 0.05) \\
        ABELL2390 &  (2.39, 0.17)  &  (1.46, 0.25)  &  --       \\
        ABELL2485 &  (1.78, 0.21)  &  (1.42, 0.44)  &  --       \\
        ABELL2537 &  (2.09, 0.26)  &  (2.60, 0.62)  &  --       \\
        ABELL2552 &  (2.12, 0.31)  &  (1.49, 0.62)  &  --       \\
        ABELL2631 &  (2.02, 0.25)  &  (0.96, 0.69)  &  (1.96, 0.09) \\
        ABELL2645 &  (1.47, 0.25)  &  (1.48, 0.45)  &  --       \\
        ABELL2813 &  (2.14, 0.21)  &  (1.79, 0.42)  &  --       \\
        RXJ1504.1$-$0248 &  (1.76, 0.24)  &  (2.97, 0.47)  &  (2.25, 0.11) \\
        RXJ1720.1+2638 &  (1.75, 0.31)  &  (1.36, 0.45)  &  (1.93, 0.06) \\
        RXJ2129.6+0005 &  (1.62, 0.30)  &  (0.71, 0.67)  &  (1.83, 0.10) \\
        ZwCl1021.0+0426 &  (1.69, 0.19)  &  (1.67, 0.38)  &  --       \\
        ZwCl1459.4+4240 &  (1.21, 0.29)  &  (2.81, 0.57)  &  --       \\
        \hline
    \end{tabular}
\end{table}

\begin{figure}
    \centering
    \includegraphics[width=\columnwidth]{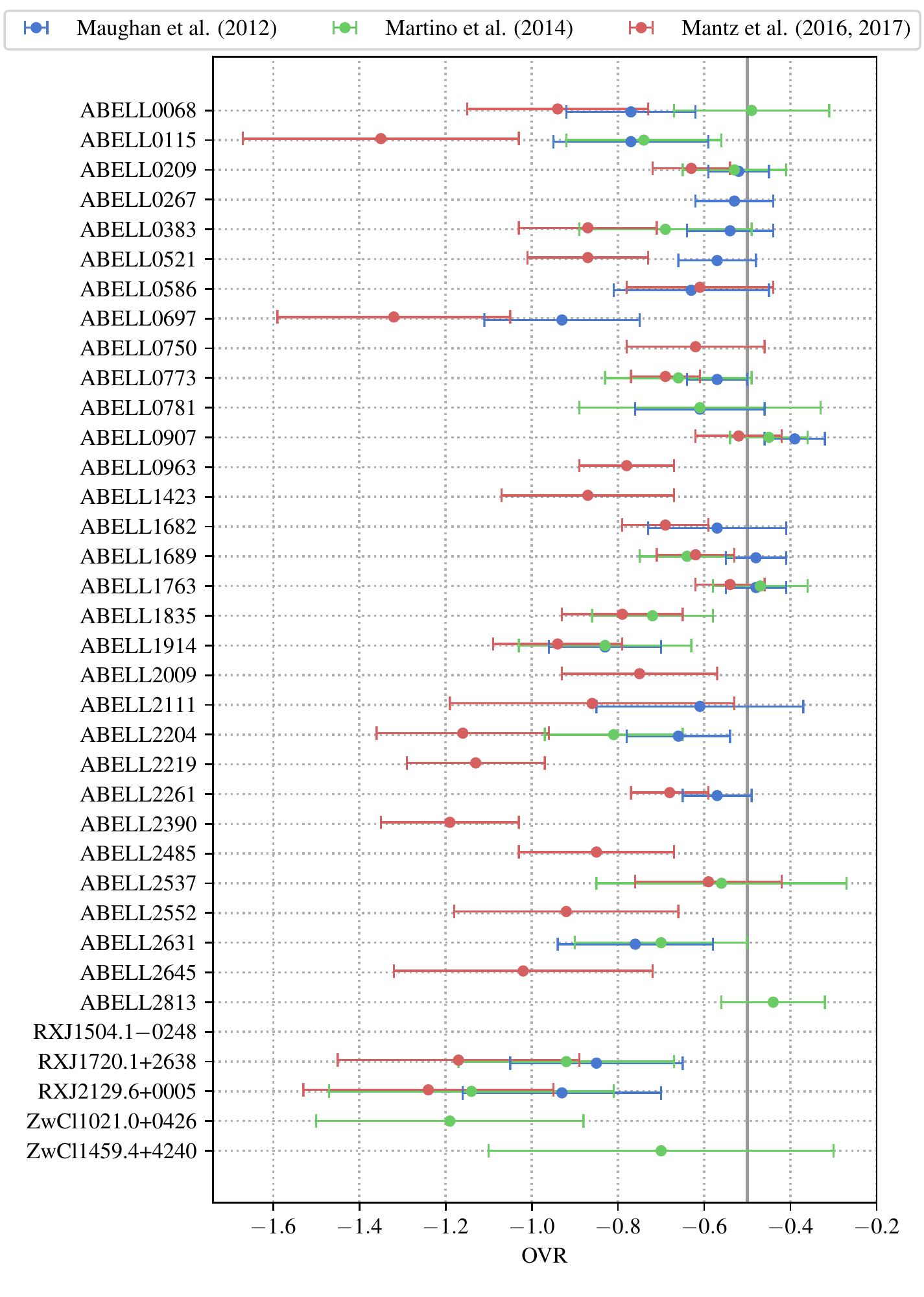}
    \caption{Processed observed virial ratios from the NFW fit parameters data
        combined X-ray temperatures from different sources. The vertical grey
        line marks the classic value $0.5$.}
    \label{fig:OVR}
\end{figure}

Since the logarithm of these distributions follow Gaussian distributions with
the usual parameters $(\mu, \sigma)$, we proceed to use linear error propagation
for the left-hand side of Eq.~(\ref{eq:LI}) with the asymmetrical measurements
$\bar x^{+\sigma_{+}}_{-\sigma_{-}}$ symmetrized to $\bar x' \pm \Delta x' =
\exp \left(\mu \pm \sigma \right)$ for the quantities $M_{200}$, $c_{200}$
and $\kT$. 
Whenever possible, the error is propagated on the combined logarithmic
quantities first, to minimize introduction of bias.
We compute the observed virial ratios following Eq.~(\ref{eq:OVR}) and present
them in Fig.~\ref{fig:OVR}.

\begin{figure}
    \centering
    \includegraphics[width=\columnwidth]{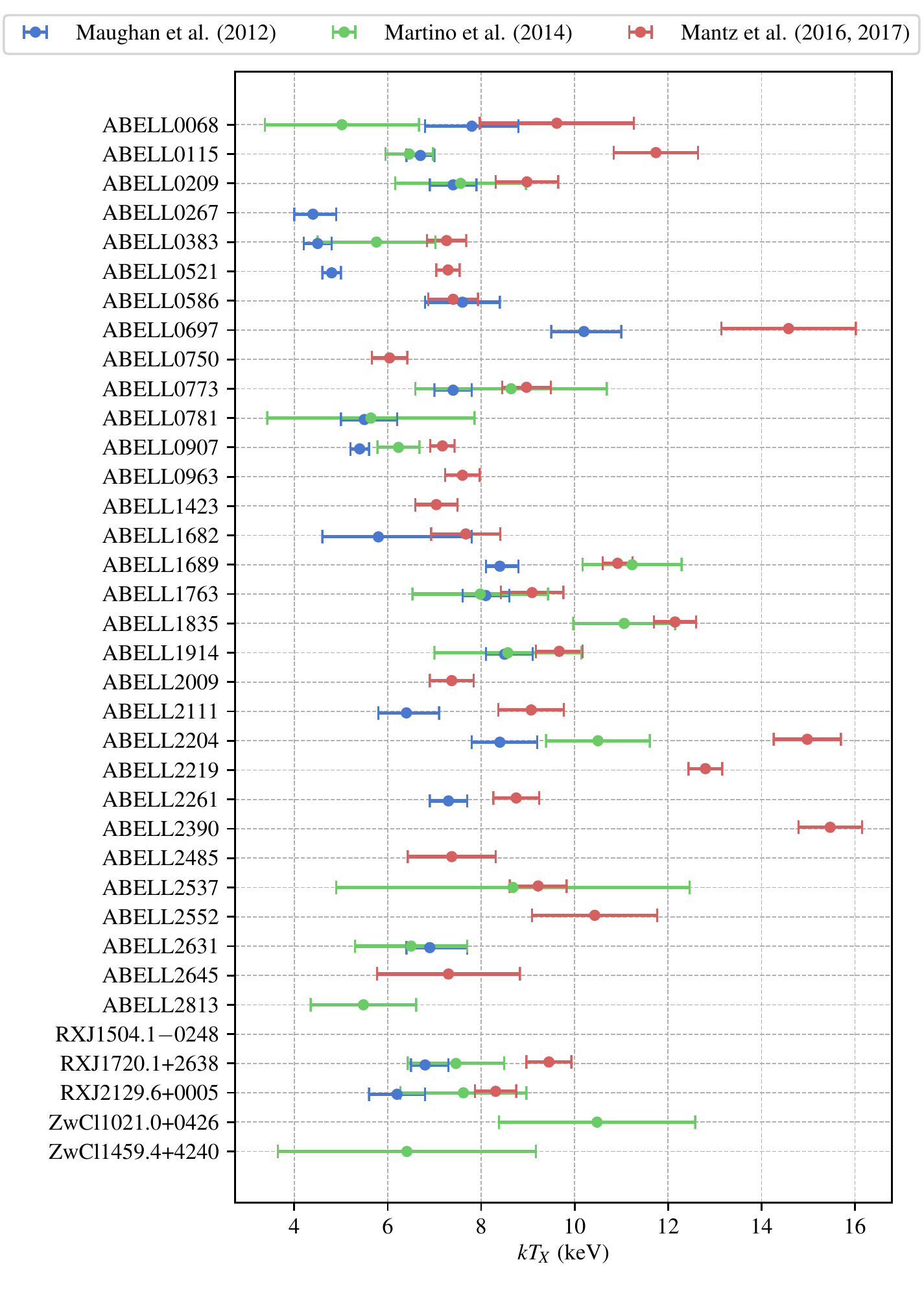}
    \caption{X-ray temperatures from different sources. Values are listed in
        Table~\ref{tab:clusters_data}.}
    \label{fig:difftemp}
\end{figure}

The differences in observed virial ratios reflect the different temperature
measurements listed in Table~\ref{tab:clusters_data} and also plotted in
Fig.~\ref{fig:difftemp}.
In view of conflicting data, we build ``gold'' samples of clusters that have at
least two temperature measurements within $1\sigma$ of each other, and consider
their average (or the average of their logarithms) for the calculations.
By inspecting Fig.~\ref{fig:difftemp}, we selected three
gold samples composed of six clusters from M12+M14:
ABELL0115, ABELL0209, ABELL0781, ABELL1763, ABELL1914,
ABELL2631; one cluster from M12+M16: ABELL0586; and four
clusters form M14+M16: ABELL0773, ABELL1689, ABELL1835 and
ABELL2537 from overlapping error bar clusters.
A sample with all eleven clusters from these samples
(referred to as GOLD) is also considered.

The Hubble function $H(z)$ in the \gls{dfe} term must be evaluated in the CDE
cosmology, thus depending on the parameters $\xi$, $h$ and $\Omega_{c0} h^2$.
In terms of these parameters, $H(z)$ is given by the Friedmann equation
(restricting to the case $w_d=-1$) in the form 
\begin{equation}
    \left[ \frac{H(z)}{100} \right]^2 = 
    h^2 + \Omega_{c0} h^2 \frac{a^{-3(1-\xi)} - 1}{1-\xi}.
\end{equation}
When $w_d$ is free, $H(z)$ is obtained from the Friedmann equation in its original form with the numerical solutions of Eqs.~(\ref{eq:Bianchi}).
We then include $H(z)$ from cosmic chronometer data \citep{moresco_6_2016}, the
JLA supernovae binned dataset \citep{betoule_improved_2014} and the local
measurement of $H_0 = \left(73.24 \pm 1.74 \right) \rmn{km} \, \rmn{s}^{-1}
\rmn{Mpc}^{-1}$ from \citet{riess_2.4_2016}, joined to the clusters data, in order
to perform the analysis outputting $h$ and $\Omega_{c0} h^2$ together with
$\xi$.

\subsection{The clusters likelihood}
\label{sec:data}
The left-hand side of equation~(\ref{eq:LI}), computed from the measurements of
mass, temperature and NFW concentration, constitutes our observable, as
explained in Sec.~\ref{ss:LIequation}.
Denoting by $f_N(x; \mu, \sigma) = (2 \uppi \sigma^2)^{-1/2} \exp
[-\left(x - \mu\right)^2/2 \sigma^2 ]$ the PDF of a Gaussian
distribution $N(\mu, \sigma)$, we assume Gaussian likelihoods
$\mathcal{L}_{\rmn{cluster}} = f_N(\rmn{EVR}; \mu, \sigma)$ for each
cluster, with $\mu$ and $\sigma$ given by the nominal value and standard
deviation of the quantity $\rmn{OVR} - \rmn{DfE}$,
to compare the predicted values of the equilibrium virial ratio 
$\rmn{EVR}(\xi) \equiv - \left(1 - 6\xi\right)/\left(2 + 3 \xi\right)$ with this
observable.

The total likelihood of a set of clusters is given by the product
$\mathcal{L}_{\rmn{clusters}} = \prod_i \mathcal{L}_{\rmn{cluster} \, i}$ of
the likelihoods of all the clusters in the given sample.
We should stress that the left-hand side of equation~(\ref{eq:LI}) depends on
the amount of matter and the Hubble parameter through $\Omega_{c0}$ and $H_0$,
or equivalently $\Omega_{c0} h^2$ and $h$, motivating us to include $H(z)$ and
supernovae data.
The parameters $\xi$ and $\gamma$ are also implicit in the likelihood
$\mathcal{L}_{\rmn{cluster}}$ through the \gls{dfe} term.
$\mathcal{L}_{\rmn{clusters}}$ is thus also multiplied by the product of the
Gaussian likelihoods $\mathcal{L}_{H(z)} = \prod_i f_N(H(z_i)_{\rmn{predicted}};
H(z_i), \sigma_{H(z_i)})$ of the $H(z)$ data and by the JLA likelihood
$\mathcal{L}_{\rmn{JLA}}$, based on estimates of binned distance modulus
obtained from the JLA supernovae sample \citep[from][]{betoule_improved_2014}: 
$\mathcal{L}_{\rmn{total}} = \mathcal{L}_{\rmn{clusters}} \times \mathcal{L}_{H(z)} \times \mathcal{L}_{\rmn{JLA}}$.
An additional nuisance parameter $\Delta M$ is included to account for a
possible shift in the absolute magnitudes of the supernovae.

We thus obtain the unnormalized posterior distribution probabilities $P(\theta \mid
D)$, for our set of parameters $\theta = \bigl\lbrace \xi, h, \Omega_{c0} h^2,
\gamma, \Delta M \bigr\rbrace$ given the data $D$ by using Bayes' theorem
\begin{equation}
    P(\theta \mid D) = \frac{\mathcal{L}_{\rmn{total}}(D \mid \theta) \, \pi(\theta)}{P(D)},
\end{equation}
where $\pi(\theta)$ is the prior probability for the parameters, assumed flat
and detailed in section \ref{sec:MCMC}.
The correct normalization of the posterior distribution is given by the
marginal likelihood or evidence $P(D)$, which is not
required for our parameter inference purposes.

\section{The MCMC analyses}
\label{sec:MCMC}
Using the EPIC code \citep{marcondes_epic_2017}, we run
\gls{mcmc} simulations for our interacting model with fixed
$w_d = -1$ using each of the four samples considered above
and with $w_d$ using the GOLD sample.
The clusters data are combined with $H(z)$ and supernovae
data in all cases.
We set flat priors over the intervals $[-0.2, 0.2]$ for
$\xi$, $[0.5, 0.9]$ for $h$, $[0.0, 0.3]$ for $\Omega_{c0}h^2$, 
$[0.00, 0.99]$ for $\gamma$, $[-1.0, 1.0]$ for $\Delta M$
and $[-2.0, -0.4]$ for $w_d$ when it is free.
The code evolved 12 independent Markov chains in each case, the convergence,
according to the Gelman-Rubin criteria for multivariate distributions
\citep{gelman_inference_1992, brooks_general_1998}, being checked with the
multivariate potential scale reduction factor $\hat R^p$ for $p$
parameters within about $5 \times 10^{-3}$ of 1.

The constraints on $\xi$ and $\rmn{EVR}$ and $w_d$ are given in
Table~\ref{tab:constraints_xi} at $1\sigma$ and $2\sigma$
confidence levels (C.L.); the other
parameters are given in Table~\ref{tab:otherparameters}.
\begin{table}
    \centering
    \caption{Constraints on the interaction strength parameter $\xi$ 
        and the derived parameter $\rmn{EVR}(\xi)$ of the CDE model from $H(z)$
        data, supernovae data and each of the clusters samples M12+M14, M12+M16,
        M14+M16 and GOLD.}
    \label{tab:constraints_xi}
    \renewcommand{\arraystretch}{1.4}
    \begin{tabular}{@{}lcccc@{}}
        \hline
        Parameter   &   Sample   & 		Best-fitting	 & 	$1\sigma$ C.L.	 & 	$2\sigma$ C.L.	\\
        \hline
        \multirow{5}{*}{$100\,\xi$} &   M12+M14   &	$-1.90$ &	$-1.86^{+1.23}_{-1.28}$	& $-1.86^{+2.62}_{-2.47}$	 \\
        {}  &   M12+M16   &	$-2.29$	&	$-1.83^{+4.65}_{-4.43}$	&	$-1.83^{+10.42}_{-8.45}$	\\
        {}  &   M14+M16	&	$-3.90$	&	$-3.94^{+1.58}_{-1.44}$	&	$-3.94^{+3.38}_{-2.81}$	\\
        {}  &   GOLD	&	$-2.85$	&	$-2.70^{+0.90}_{-0.91}$	&	$-2.70^{+1.82}_{-1.81}$	\\
        {}  &   GOLD ($w_d$ free)   &   $-3.05$	&	$-2.79^{+0.94}_{-0.93}$	&	$-2.79^{+1.97}_{-1.79}$	\\
        \hline
        \multirow{5}{*}{$\rmn{EVR}(\xi)$}   &   M12+M14 &   $-0.57$	&	$-0.57 \pm 0.05$	&	$-0.57 \pm 0.10$ \\
        {}  &   M12+M16 &   $-0.59$	&	$-0.55^{+0.17}_{-0.19}$	& $-0.55^{+0.36}_{-0.37}$ \\
        {}  &   M14+M16 &   $-0.66$	&	$-0.66^{+0.07}_{-0.06}$	&	$-0.66^{+0.14}_{-0.12}$	\\
        {}  &   GOLD    &   $-0.61$	&	$-0.61^{+0.04}_{-0.03}$	& $-0.61^{+0.08}_{-0.07}$ \\
        {}  &   GOLD ($w_d$ free)   &   $-0.62$	&	$-0.61 \pm 0.04$	&   $-0.61^{+0.08}_{-0.07}$ \\
        \hline
        $w_d$	&	GOLD ($w_d$ free)   &	$-0.98$	&	$-0.99^{+0.16}_{-0.14}$	&	$-0.99^{+0.29}_{-0.32}$	\\
        \hline
    \end{tabular}
\end{table}

\begin{figure}
    \centering
    \subfloat{\includegraphics[width=0.47\columnwidth]{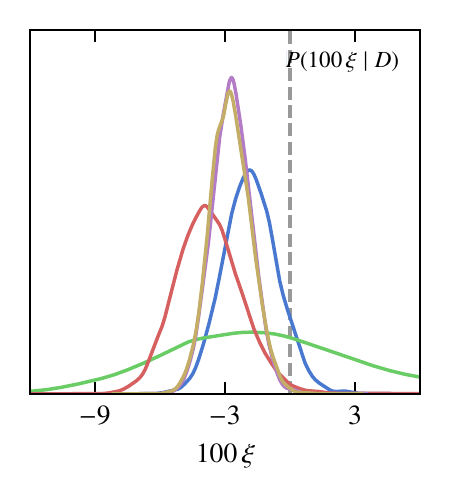}} \, \quad
    \subfloat{\includegraphics[width=0.47\columnwidth]{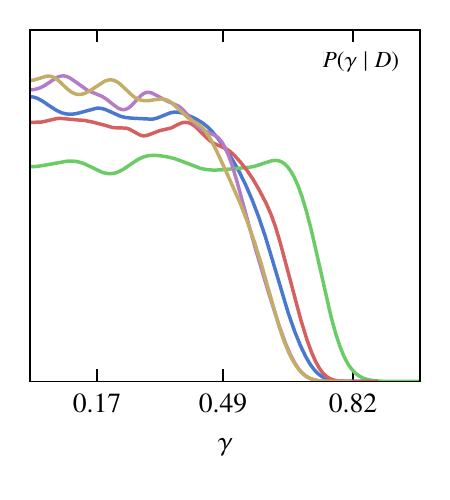}} \\
    \subfloat{\includegraphics[width=0.47\columnwidth]{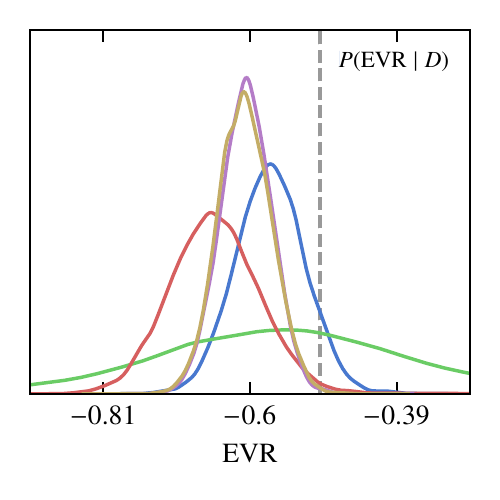}} \, \quad
    \subfloat{\includegraphics[width=0.47\columnwidth]{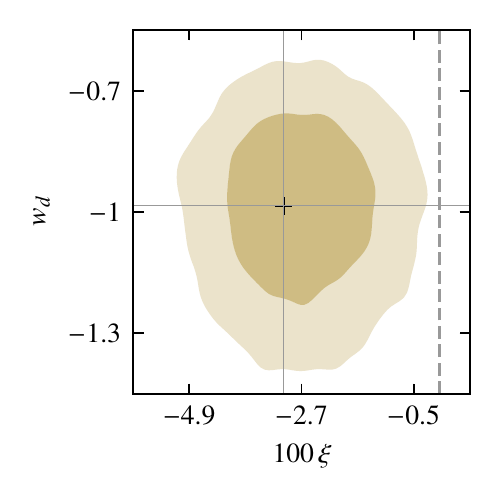}}
    \caption{Marginalized distributions of $\xi$, $\gamma$ and $\rmn{EVR}(\xi)$
        for the samples M12+M14 (blue), M12+M16 (green),
        M14+M16 (red), GOLD with $w_d$ fixed (violet) and
        GOLD with $w_d$ free (brown). 
        Dashed grey lines mark the no-interaction values
        $\xi = 0$ and $\rmn{EVR} = -0.5$.
        The last panel shows the two-dimensional joint-posterior distribution of
        the parameters $\xi$ and $w_d$ when this parameter
        is free, using the sample GOLD.}
    \label{fig:marg_dist}
\end{figure}

In Fig.~\ref{fig:marg_dist} we plot the marginalized distributions
of the parameters $\xi$, $\gamma$ and the joint-posterior distribution of $\xi$
and $w_d$ when this is also free, with the sample GOLD.
We note that the analysis with $w_d$ free has no effect on the marginalized
distribution of the interaction constant (the corresponding
violet and brown curves are almost indistinguishable),
although it does affect the distribution of $\Omega_{c0}
h^2$ (not plotted).

\section{Results}\label{sec:Results}
Constraints with sample M12+M16 are compatible with $\xi = 0$ within $1\sigma$,
while M12+M14, M14+M16, GOLD and GOLD with $w_d$ free give
$1.44\sigma$, $2.30\sigma$, $2.80\sigma$ and $2.77\sigma$
detections, respectively.
When we let the dark energy equation-of-state parameter
vary, it can be noted from the joint-posterior distribution
that $w_d$ and $\xi$ are not correlated, hence the
constraints on $\xi$ (and also on all other parameters except
$\Omega_{c0} h^2$) are practically unchanged.
This can be seen in the lack of strong difference in
Fig.~\ref{fig:marg_dist} between the two GOLD
distributions.

If we disregard the M12+M16 sample, as it  only contains one
cluster, it appears that we have $2\sigma$ to 3$\sigma$ detection of
the DE-DM interaction, a slight improvement on previous
results of the virial detection idea
\citep[e.g.][]{bertolami_dark_2007}.  As discussed
previously, the main problem appears from the inconsistent
X-ray temperature detections, with no present guiding
principle to favour one dataset over another. We turned the
difficulty by selecting in the three datasets available to
us that contained consistent clusters and compiled them in a
GOLD sample. The method clearly improves detection when
stacking as many clusters as possible: the distributions for
$\xi$ and $\rmn{EVR}$ on Fig.~\ref{fig:marg_dist} are more
peaked for larger samples.

\section{Discussion and conclusions}\label{sec:conclusions}
In this paper we have continued the works of
\cite{bertolami_dark_2007, le_delliou_dark_2007,
    bertolami_dark_2008, bertolami_abell_2009,
    bertolami_testing_2012, abdalla_signature_2009,
    abdalla_signature_2010,
    he_imprint_2010,le_delliou_non-virialized_2015} on
virial detection of dark sector interaction. The approach of
\cite{le_delliou_non-virialized_2015} for non-virialised
clusters was improved to obtain consistent results. Based on
evaluation of the dynamical out-of-equilibrium state
independent of the details of each cluster's astrophysical
history, the method relies on a set of simplifying
reasonable assumptions. Although the convergence ansatz
could be debatable, its general features prove to provide
enough power to the method so as to be able to yield
consistent results. From a sample of 50 clusters with full
necessary data, consistency led us to trim down to a maximum
of 11 clusters. The results range from no detection, but for
a single cluster sample, to 3$\sigma$ detection, with
improvement when the samples are larger. This is a strong
indication that the method is sound and likely to yield a
clear answer to dark sector interaction question, given
larger samples of clusters, with clear guidance on the X-ray
temperature detection reliability and robust weak lensing
determination.

This is why the detection of interaction in the dark sector
(or its ruling out) will greatly benefit from future
instruments and surveys.
In particular, increasing the number of clusters
with mass distribution measurements through lensing effects
(which need deep imaging and large field-of-view) with the
next generation of telescopes, such as the Thirty Meter
Telescope \citep[TMT,][]{skidmore_thirty_2015}, 
the Giant Magellan Telescope \citep[GMT,][]{johns_giant_2012}
and the European Extremely Large Telescope \citep[E-ELT,][]{mcpherson_e-elt_2012}.
Likewise, the X-ray detected clusters will increase in the
next few years with the extended ROentgen Survey with an
Imaging Telescope Array \citep[\textit{eROSITA},][]{merloni_erosita_2012}.
With these perspectives in observations and the method
finalised here, we are confident that a reliable dark sector
interaction detection is within reach.

\section{Acknowledgements}

We thank Nobushiro Okabe for generously sharing with us his NFW fits of the LoCuSS clusters. 
The work of M.Le D.~has been supported by Lanzhou University
starting fund and PNPD/CAPES20132029. M.Le D.~also wishes to
acknowledge DFMA/IF/USP and IFT/UNESP where this work was
initiated. G.B.L.N.~thanks financial support form CNPq and FAPESP (2018/17543-0).
%%%%%%%%%%%%%%%%%%%%%%%%%%%%%%%%%%%%%%%%%%%%%%%%%%



\bibliographystyle{mnras}
\bibliography{refs} % if your bibtex file is called example.bib

%%%%%%%%%%%%%%%%%%%%%%%%%%%%%%%%%%%%%%%%%%%%%%%%%%

%%%%%%%%%%%%%%%%% APPENDICES %%%%%%%%%%%%%%%%%%%%%

%%%%%%%%%%%%%%%%%%%%%%%%%%%%%%%%%%%%%%%%%%%%%%%%%%
\appendix

\section{Constraints on the other parameters}

Completing Table~\ref{tab:constraints_xi}, we present here in Table~\ref{tab:otherparameters} the
remaining constraints on the other parameters of our
analyses.
The confidence intervals reported for $\gamma$ without
central values reflect the fact that its distributions are
poorly constrained, only suppressed by the singularity as
$\gamma$ approaches 1 but otherwise flat.
The exact values at which the distributions become suppressed
are sensitive to our arbitrary choice of $t_0 = 1 \,
\rmn{km}^{-1} \, \rmn{s} \, \rmn{Mpc}$ (see
Sec.~\ref{sss:DfE}). However, this does not affect our
results, which are based on marginalizing this parameter over
all values allowed by our priors.
Constraints on the parameters $\Omega_{c0}$ and
$\Omega_{d0}$ derived from $\Omega_{c0}h^2$ and $h$ are also
listed.

\begin{table}
    \centering
    \caption{Constraints on parameters of CDE model from $H(z)$ data, supernovae
        data and each of the clusters samples M12+M14, M12+M16, M14+M16 and
        GOLD.}
    \label{tab:otherparameters}
    \renewcommand{\arraystretch}{1.4}
    \begin{tabular}{@{}lcccc@{}}
        \hline
        Parameter	 &  Group   & 		Best-fitting	 & 	$1\sigma$ C.L.	 & 	$2\sigma$ C.L.	\\
        \hline
        \multirow{5}{*}{$h$}      &	M12+M14   &	$0.72$	&	$0.72^{+0.01}_{-0.02}$	&	$0.72^{+0.02}_{-0.03}$	\\
        {}       &   M12+M16   &	$0.72$	&	$0.71^{+0.02}_{-0.01}$	&	$0.71^{+0.03}_{-0.02}$	\\
        {}      &	M14+M16	&	$0.71$	&	$0.72^{+0.01}_{-0.02}$	&	$0.72^{+0.02}_{-0.03}$	\\
        {}		&	GOLD	&	$0.71$	&	$0.71^{+0.02}_{-0.01}$	&	$0.71^{+0.03}_{-0.02}$	\\
        {}  &   GOLD ($w_d$ free)   &   $0.71$	& $0.71^{+0.02}_{-0.01}$	&	$0.71^{+0.03}_{-0.02}$ \\
        \hline
        \multirow{5}{*}{$\Omega_{c0}h^2$}      &	M12+M14   &	$0.13$	&   $0.13^{+0.02}_{-0.01}$	&	$0.13^{+0.03}_{-0.02}$	\\
        {}    &   M12+M16   &	$0.13$	&	$0.13^{+0.03}_{-0.02}$	&	$0.13^{+0.06}_{-0.03}$	\\
        {}      &	M14+M16	&	$0.13$	&	$0.13 \pm 0.01$	&	$0.13 \pm 0.02$	\\
        {}      &	GOLD    &	$0.13$	&	$0.13 \pm 0.01$	&	$0.13 \pm 0.02$	\\
        {}      &	GOLD ($w_d$ free)   &	$0.14$	&	$0.15^{+0.02}_{-0.03}$	&	$0.15^{+0.04}_{-0.07}$	\\
        \hline
        \multirow{5}{*}{$\gamma$}    &   M12+M14   & 	$0.03$	 & 	$0 < \gamma \leqslant 0.41$  &   $0 < \gamma \leqslant 0.61$	   \\
        {} &   M12+M16   & 	$0.20$	 & 	$0 < \gamma \leqslant 0.64$ &   $0 < \gamma \leqslant 0.71$	   \\
        {}  &   M14+M16   & 	$0.10$	 & 	$0 < \gamma \leqslant 0.43$  &   $0 < \gamma \leqslant 0.63$  \\
        {}  &   GOLD   & 	$0.01$	 & 	$0 < \gamma \leqslant 0.39$  &   $0 < \gamma \leqslant 0.57$  \\
        {}  &   GOLD ($w_d$ free)   & 	$0.11$	&	$0 < \gamma \leqslant 0.38$	&	$0 < \gamma \leqslant 0.57$  \\
        \hline
        \multirow{5}{*}{$\Delta M$}      &	M12+M14   &	$0.04$	& $0.04 \pm 0.04$	&	$0.04^{+0.08}_{-0.09}$	\\
        {}  &   M12+M16   &	$0.04$	&	$0.04 \pm 0.04$	&	$0.04^{+0.08}_{-0.09}$	\\
        {}  &	M14+M16   &	$0.03$	&	$0.03^{+0.05}_{-0.04}$	&	$0.03^{+0.09}_{-0.08}$	\\
        {}  &	GOLD  &	$0.03$	&	$0.04 \pm 0.04$	&	$0.04^{+0.08}_{-0.09}$	\\
        {}  &	GOLD ($w_d$ free)  &	$0.04$	&	$0.04 \pm 0.04$	&	$0.04 \pm 0.09$	\\
        \hline
        \multirow{5}{*}{$\Omega_{c0}$}	&   M12+M14	&	$0.26$	& $0.26^{+0.03}_{-0.02}$	&	$0.26^{+0.06}_{-0.05}$ \\
        {}   &   M12+M16   &	$0.26$	&	$0.26^{+0.05}_{-0.04}$	&	$0.26^{+0.12}_{-0.07}$	\\
        {}  &	M14+M16   &	$0.25$	&	$0.25^{+0.02}_{-0.03}$	&	$0.25 \pm 0.05$	\\
        {}  &	GOLD   &	$0.25$	&	$0.26^{+0.02}_{-0.03}$	&	$0.26^{+0.04}_{-0.05}$	\\
        {}  &	GOLD ($w_d$ free)   &	$0.28$	&	$0.29^{+0.04}_{-0.06}$	&	$0.29^{+0.09}_{-0.13}$	\\
        \hline
        \multirow{5}{*}{$\Omega_{d0}$}	&	M12+M14	&	$0.74$	& $0.74^{+0.02}_{-0.03}$	&	$0.74^{+0.05}_{-0.06}$ \\
        {}   &   M12+M16   &	$0.74$	&	$0.74^{+0.04}_{-0.05}$	&	$0.74^{+0.07}_{-0.12}$	\\
        {} &	M14+M16   &	$0.75$	&	$0.75^{+0.03}_{-0.02}$ 	&	$0.75 \pm 0.05$	\\
        {} &	GOLD  &	$0.75$	&	$0.74^{+0.03}_{-0.02}$ 	&	$0.74^{+0.05}_{-0.04}$	\\
        {} &	GOLD ($w_d$ free)  &	$0.72$	&	$0.71^{+0.06}_{-0.04}$ 	&	$0.71^{+0.13}_{-0.09}$	\\
        \hline
    \end{tabular}
\end{table}

% Don't change these lines
\bsp	% typesetting comment
\label{lastpage}
\end{document}